\documentclass[aps,prl,amsmath,amssymb,twocolumn,nobibnotes,nofootinbib, showpacs, showkeys, nofootinbib]{revtex4}
\usepackage{pslatex,graphicx,dcolumn,bm,natbib}
\usepackage{amsmath}
\usepackage{multirow}
\usepackage{subfigure}
\usepackage{color}
\usepackage[colorlinks,linkcolor=blue,anchorcolor=blue,citecolor=blue]{hyperref}

\setcitestyle{super}
\date{\today}

\begin{document}

\title{Level statistics and Anderson delocalization in two-dimensional granular materials}


\author{Ling Zhang$^{1,2}\dagger$, Yinqiao Wang$^2\dagger$, Jie Zheng$^2$, Aile Sun$^2$, Xulai Sun$^2$, Yujie Wang$^2$, Walter Schirmacher$^{3}$, and Jie Zhang$^{2,4,\ast}$}
\affiliation{$^1$School of Automation, Central South University, Changsha 410083, China.}
\affiliation{$^2$School of Physics and Astronomy, Shanghai Jiao Tong University, Shanghai 200240, China}
\affiliation{$^3$Institut f$\ddot{u}$r Physik, Universit$\ddot{a}$t Mainz, Staudinger Weg 7, D-55099 Mainz, Germany}
\affiliation{$^4$Institute of Natural Sciences, Shanghai Jiao Tong University, Shanghai 200240, China}

\begin{abstract}
	Contrary to the theoretical predictions that all waves in two-dimensional disordered materials are localized, Anderson localization is observed only for sufficiently high frequencies in an isotropically jammed two-dimensional disordered granular packing of photoelastic disks. More specifically, we have performed an experiment in analyzing the level statistics of normal mode vibrations. We observe delocalized modes in the low-frequency boson-peak regime and localized modes in the high frequency regime with the crossover frequency just below the Debye frequency. 
We find that the level-distance distribution obeys Gaussian-Orthogonal-Ensemble (GOE) statistics, i.e. Wigner-Dyson distribution, in the boson-peak regime, whereas those in the high-frequency regime Poisson statistics is observed. The scenario is found to coincide with that of harmonic vibrational excitations in three-dimensional disordered solids.
\end{abstract}

-

\maketitle

Wave localization in amorphous materials has become one of the most spectacular phenomena in the field of condensed-matter physics since it was proposed by Anderson to describe the localization of electron waves \cite{1958-PW.Anderson-PhyscalReview, 1979-E.A.brahams-P.W.Anderson-PRL}. With the continuous experimental advancement,  Anderson localization has been found ubiquitously, such as the localization of ultra-cold atoms\cite{2008-J.Billy-Nature,2008-G.Roati-Nature,2011-SS.Kondov-Science, 2015-M.Schreiber-Science} and light and sound\cite{2000-AA.Chabanov-Nature,2007-TSchwartz-Nature,2008-HefeiHu-NatPhys,2018-Schirmacher-PRL}.
One commonly accepted wisdom is that Anderson localization depends on dimension: waves are in general localized in one and two dimensions \cite{1985-Lee-RMP}, including acoustic waves \cite{1983-John-PRB,1985-Kirkpatrick-PRB,2010-Monthus-PRB}. Only in three dimensions a mobility edge separating localized from extended waves is expected. It has, however been pointed out that for weak disorder \cite{1984-Economou-PRB,1983-John-PRB,2010-Monthus-PRB} the localization length may become macroscopically large. 

A number of studies \cite{1985-E.Akkermans-R.Maynard-PRB,1989-PB.Allen-PRL,1993-PB.Allen-PRB} tried to relate Anderson localization to anomalous vibrational properties of phonons in glasses, in particular the so-called boson peak, which is an enhancement of the vibrational density of states, compared with Debye's $\omega^2$ law \cite{1986-JJ.Freeman-AC.Anderson-PRB,1988-Y.Leggett-CommCondMatPhys,1998-W.Schirmacher-PRL,2001-SN.Taraskin-PRL,2013-W.Schirmacher-PSSB,Baggioli2019vibrational,Baggioli2019universal, 2005-LE.Silbert-PRL,2008-H.Tanaka-NatureMat,2009-G.Monaco-PNAS,2013-W.Schirmacher-ScientificRep,2017-Mizuno-Ikeda-PNAS,Buchenau-PRB-1986,1986-AP.Sokolov-SolidStateCom,2010-Douchot-SoftMatter,2011-Chumakov-PRL,2014-Chumakov-PRL,2017-LZ-JZhang-NatCommun,Wang2018-disentangling,2020-Wang-PRL}. 
Skepticism about this interpretation has been raised in later theoretical and numerical studies in three-dimensional systems \cite{1994-P.Sheng-PRL,1998-W.Schirmacher-PRL, 1999-W.Schirmacher-AnnPhys}, which point out that the boson peak and Anderson localization are two separate identities, obeying different statistics.
Moreover, the boson peak is a universal characteristic of amorphous materials, independent of the actual dimensionality as observed in  numerical \cite{2003-OHern-PRE, 2005-LE.Silbert-PRL, 2014-M.Wyart-SoftMatter,2015-Franz-PNAS,2016-Patrick-PRL, 2008-H.Tanaka-NatureMat, 2017-Mizuno-Ikeda-PNAS} and experimental \cite{2011-ChenKe-PRL,2017-LZ-JZhang-NatCommun, Wang2018-disentangling} studies of two-dimensional systems.
Therefore, the situation is not clear. One essential scientific question is whether the vibrational properties of phonons in disordered systems are truly localized in two dimensions as commonly believed \cite{1985-Lee-RMP,1985-Kirkpatrick-PRB}. 
This is the main focus of the present work.

To address the above important question, we analyze the level statistics, which is extremely powerful in distinguishing the regime of Anderson localization from the delocalized one \cite{1990-Izrailev-Phys-Rep}. Schirmacher et al \cite{1998-W.Schirmacher-PRL} found in a model calculation of a cubic lattice with disordered nearest-leighbor force constants that  
the level distances obey Gaussian-Orthogonal-Ensemble (GOE) statistics in the boson-peak regime and Poisson statistics only near the upper band edge,
well above the boson-peak regime. They concluded, in accord with earlier
work \cite{1993-W.Schirmacher-SolidStateCommun}, that the modes
around the boson peak are delocalized and localized states exist
only at high frequencies.
Recently it was discovered that a lattice of coupled masses and springs with random values \cite{2012-W.Schirmacher-EPL,2012-W.Schirmacher-J.PhysCondensMatter} and truncated Lennard-Jones fluid \cite{2009-BJ.Huang-TM.Wu-PRE} exhibit Anderson universality of disordered phonons, which means that the localization transition is in the same universality class as the electronic Anderson model. In both investigations the localization frequency is located near the upper band edge. 

In a related, albeit different context, level-distance statistics is closely related to the thermalisation of an isolated quantum many-body system, in which the GOE statistics represents the thermalizing ergodic phase and the Poisson statistics represent the nonthermalizing phase \cite{2016-M.Serbyn-PRB}. From a dynamical perspective, level-distance statistics is deeply connected with the picture of Brownian motion in the space of random Hamiltonian matrices \cite{1962-Dyson-JMP, 1996-Chalker-PRL}.

It is of great challenge to directly investigate the level-distance statistics experimentally. Neutron scattering is not appropriate in resolving the level distance of vibrational energies due to its finite energy resolutions \cite{Mitchell2005-InelasticBook}. The similar limitation exists in X-ray and Raman scattering. Hence, measuring level distance in molecular glasses is extremely difficult. 
Alternatively, one may turn to experiments with colloids, however, there difficulties may arise with the covariance matrix method \cite{Sussman2015-Strain}. 

In recent years, Zhang et al \cite{2017-LZ-JZhang-NatCommun} and Wang et al \cite{Wang2018-disentangling} have successfully used isotropically jammed disordered macroscopic packings of photoelastic disks to perform experimental studies of vibrations in disordered systems and in particular
the nature of the boson peak. In these studies\cite{2017-LZ-JZhang-NatCommun,Wang2018-disentangling}, the dynamical matrix can be directly constructed from the experimental measurements of the particle positions and the contact forces between the particles, which permits obtaining the whole set of eigenvalues of the vibrational modes, opening the possibility of investigating the level-distance statistics of vibrational modes of jammed disordered granular matter. 

In this paper, we analyze the level-distance statistics of the vibrational normal modes and find compelling experimental evidence supporting that Anderson localization exists at sufficiently high frequencies in a two-dimensional isotropically jammed disk packing. We further identify a localization-delocalization  transition slightly below Debye frequency. All states below this mobility edge are delocalized.
We find that the mobility edge corresponds to a localization length that is comparable to the system size. For smaller frequencies the localization length is much larger than the system size due to the Rayleigh law for the mean-free path. This observation is consistent with the theoretical description of the localization length as a rapid growth function of the decrease of frequency in two dimension\cite{1983-John-PRB,1985-Lee-RMP,2018-Schirmacher-PRL}.

More specifically, using the photoelastic technique \cite{2005-Trush-Nature,2017-LZ-JZhang-NatCommun,Wang2018-disentangling}, we directly measure the contact forces and contact positions between the disks in addition to the positions of the disk centers, which allows us to construct the Hessian matrix to obtain a complete set of the eigenvalues of the normal modes \cite{2017-LZ-JZhang-NatCommun,Wang2018-disentangling}.
By analyzing the level-distance statistics of the eigenvalues, we observe that the distribution of the level distance transforms from the GOE to the Poisson statistics, when the frequency increases from the boson-peak regime to higher frequencies. The appearance of the Poisson statistics indentifies the Anderson-localization regime. The crossover frequency from delocalization to localization occurs at a frequency $\sim12\%$ below the Debye frequency $\omega_D$ \cite{2017-LZ-JZhang-NatCommun, Wang2018-disentangling},i.e. much above the boson peak. As mentioned above, the level-distance statistics in the  boson-peak regime obeys the Wigner-Dyson statistics (i.e., the GOE statistics). In addition, the two regimes are well separated from each other, as confirmed by a finite-size analysis: the boson-peak regime is within the frequency range of $\omega \le 0.57\omega_D$ with the peak frequency $\sim0.13\omega_D$; the Anderson-localization regime is within $\omega \ge \omega_c$, in which $\omega_c\approx 0.88\omega_D$. These findings are very similar to what is found in three-dimensional model systems
\cite{1998-W.Schirmacher-PRL,1993-W.Schirmacher-SolidStateCommun,2012-W.Schirmacher-EPL,2012-W.Schirmacher-J.PhysCondensMatter,2009-BJ.Huang-TM.Wu-PRE}.

We prepared disordered isotropically jammed granular packings using a biaxial apparatus, which consists of a square frame of four mobile walls mounted on a powder-lubricated glass plate. Right under the plate, there is a circular polarizer sheet, below which a uniform LED light source is mounted.  A high-resolution camera is placed two meters above the center of the biaxial apparatus, and a mobile circular polarizer sheet is placed rightly below the camera to record images of particle configurations and stresses. To create disordered packings, we filled the square frame with $\sim 2500$ bi-disperse photoelastic disks of sizes of 1.4 cm and 1.0 cm and a number ratio of 1:1 to maximize the degree of disorder. First, we prepared a random packing slightly below the jamming point \cite{2017-LZ-JZhang-NatCommun,Wang2018-disentangling}. Next, we applied an isotropic compression to the random initial packing until its packing fraction reached the jamming point $\sim84\%$. At this point, we applied a homogeneous tapping to create an isotropic and stress-free packing, which is equivalent to a corresponding packing of frictionless particles near the jamming point\cite{2003-OHern-PRE}. Then, we continued to compress the system quasi-statically and isotropically to prepare jammed packings at different pressure. Moreover, with an accurate measurement of disk positions and interaction forces between disks, we then applied the pre-calibrated curves of contact forces (i.e., the normal and tangential components) versus deformation and determined the spring constants at each contact point. Finally, we mapped the isotropically jammed disk packing to a disordered network of coupled masses and springs, which is similar to the coupled harmonic oscillators  of random spring constants except the disordered network structure \cite{1998-W.Schirmacher-PRL,2012-W.Schirmacher-EPL}. 

\begin{figure}
\centerline{\includegraphics[trim=0cm 0cm 0cm 0cm, width=1\linewidth]{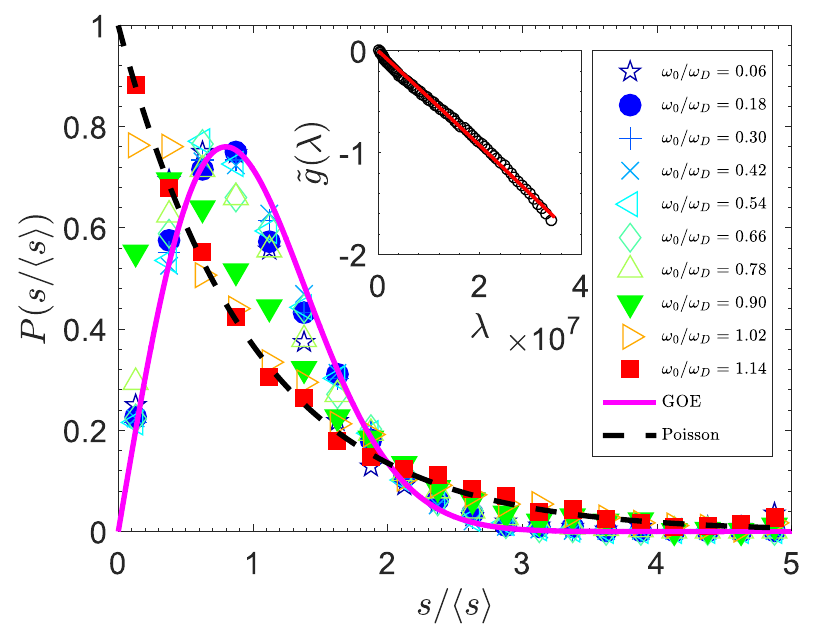}}
\caption{\label{fig:figure1}
	Main panel: Statistical distributions $p(s/\langle s \rangle)$ of the level distance $s$ normalized by its mean value $\langle s \rangle$ for 10 equal frequency intervals $\Delta \omega/\omega_D=0.12$, where $\omega_D$ is the Debye frequency. Here, $\omega_0/\omega_D$ is the re-scaled center frequency of a given interval. Lines of two different colors represent: (1) Gaussian orthogonal ensemble(GOE) statistics (a magenta solid line, i.e., $p(s/\langle s \rangle)=\frac{\pi}{2\langle s \rangle}se^{-\frac{\pi (s/\langle s \rangle)^2}{4}}$); (2) Poisson statistics (a black dashed line, i.e.,$p(s/\langle s \rangle)=e^{-s/\langle s \rangle}$). Note that all distributions $p(s/\langle s \rangle)$ are normalized to 1. Here, all results are ensemble averaged over 9 independent realizations of the same pressure $P=26.5N/m$. Results of other pressure are similar, and error bars are within symbol sizes.}

	Inset: Phase parameter $\tilde{g}(\lambda)$ is defined as $\tilde{g}(\lambda)\equiv log(1-F_{\lambda}(\lambda))$, in which $F_{\lambda}(\lambda)$ is the cumulative distribution function of $\lambda$ and $\lambda$ is the eigenvalue of the Hessian matrix of a given packing and $\lambda \equiv \omega^2$. The red solid line is a quadratic fitting function, and the exponential of such function yields $\tilde{F_{\lambda}}(\lambda)$, from which we define the level distance $s$ of two adjacent levels of $i$ and $i+1$ as $s\equiv |\epsilon_{i+1}-\epsilon_i|$ and $\epsilon_i=\tilde{F_{\lambda}}(\lambda_i)$ \cite{1998-W.Schirmacher-PRL,1999-W.Schirmacher-AnnPhys}.
\end{figure}

We construct a Hessian matrix following a standard procedure \cite{Lemaitre-Maloney-PRE-2006,EllenbroekPRE2009,2014-MWyart-PNAS, 2017-LZ-JZhang-NatCommun,Wang2018-disentangling}, whose eigenvalues are $\lambda=\omega^2$, with $\omega$ being the angular frequencies. 

First, we calculate the cumulative distribution function(CDF)
$F_{\lambda}(\lambda)
=\int g_{\lambda}(\lambda)d\lambda
=\int g_{\omega}(\omega)\frac{d\lambda}{2\omega}
=\int g_{\omega}(\omega)d\omega
=F_{\omega}(\omega)$
in which the $g_{\omega, \lambda}(\omega)$ denotes the density of states of $\omega$ or $\lambda$. Therefore, we just need to calculate $F_{\omega}(\omega)$. To smooth $F_{\lambda}(\lambda)$, we define a phase parameter $\tilde{g}(\lambda)\equiv log(1-F_{\lambda}(\lambda))$, as shown in the inset of Fig.~\ref{fig:figure1}. Next, following spectrum unfolding \cite{1998-W.Schirmacher-PRL,2001-S.Sastry-PRE}, we first use a quadratic form $A\lambda^2+B\lambda$ to fit those data points: $\tilde{g}(\lambda)= -3.95\times 10^{-17} \lambda^2-4.57\times10^{-8}\lambda$, and, then, we obtain a smooth curve of $\tilde{F}_{\lambda}(\lambda)=1-e^{A\lambda^2+B\lambda}$. Finally, we obtain statistical distributions of the nearest-neighbor distance $s$ normalized by its mean $\langle s \rangle$ within different frequency intervals \cite{1998-W.Schirmacher-PRL,1999-W.Schirmacher-AnnPhys,1994-SN.Evangelou-PRB,1997-B.Kramer-PRL,1999-P.Carpena-PRB,2001-S.Sastry-PRE,2004-C.Stefano-PRE,2009-BJ.Huang-TM.Wu-PRE,2016-M.Serbyn-PRB},
where $s\equiv |\epsilon_{i+1}-\epsilon_i|$ and $\epsilon_i=\tilde{F_{\lambda}}(\lambda_i)$ \cite{1998-W.Schirmacher-PRL,1999-W.Schirmacher-AnnPhys}. The distributions $p(s/\langle s \rangle)$ are plotted in Fig.~\ref{fig:figure1}, in which the center frequencies $\omega_0$ of  intervals,  normalized by the Debye frequency $\omega_D$, are listed in the legend. Here, only results at pressure $p=26.5N/m$ are shown. The results at other pressures are similar. The solid lines of two different colors represent two different types of statistics, as shown in Fig.~\ref{fig:figure1}: (1) GOE statistics (the magenta solid line), which are defined as $p(s/\langle s \rangle)=\frac{\pi}{2}s/\langle s \rangle e^{-\frac{\pi {(s/\langle s \rangle)}^2}{4}}$\cite{1998-W.Schirmacher-PRL,1999-W.Schirmacher-AnnPhys,2009-BJ.Huang-TM.Wu-PRE}; (2) Poisson statistics (the black dashed line), which are defined as $p(s/\langle s \rangle)=e^{-s/\langle s \rangle}$ \cite{1993-BI.Shklovskii-PRB,1998-W.Schirmacher-PRL,1997-B.Kramer-PRL,2009-BJ.Huang-TM.Wu-PRE}. When $\omega_0/\omega_D < 0.88 $ rad/s, the $p(s)$ are well fitted with the GOE statistics, which manifests level repulsion due to the random-matrix symmetry\cite{1998-W.Schirmacher-PRL,2016-M.Serbyn-PRB,1962-Dyson-JMP,1996-Chalker-PRL}, whereas for $\omega_0/\omega_D > 0.88$, the $p(s/\langle s \rangle)$ is better described by the Poisson statistics, signifying Anderson localization \cite{1993-BI.Shklovskii-PRB,1998-W.Schirmacher-PRL,1997-B.Kramer-PRL,2009-BJ.Huang-TM.Wu-PRE}, as shown in Fig.~\ref{fig:figure1}. The crossover $\omega_c$ 
is between $0.8\omega_D$ and $0.9\omega_D$.

\begin{figure}
\centerline{\includegraphics[trim=0cm 0cm 0cm 0cm, width=1\linewidth]{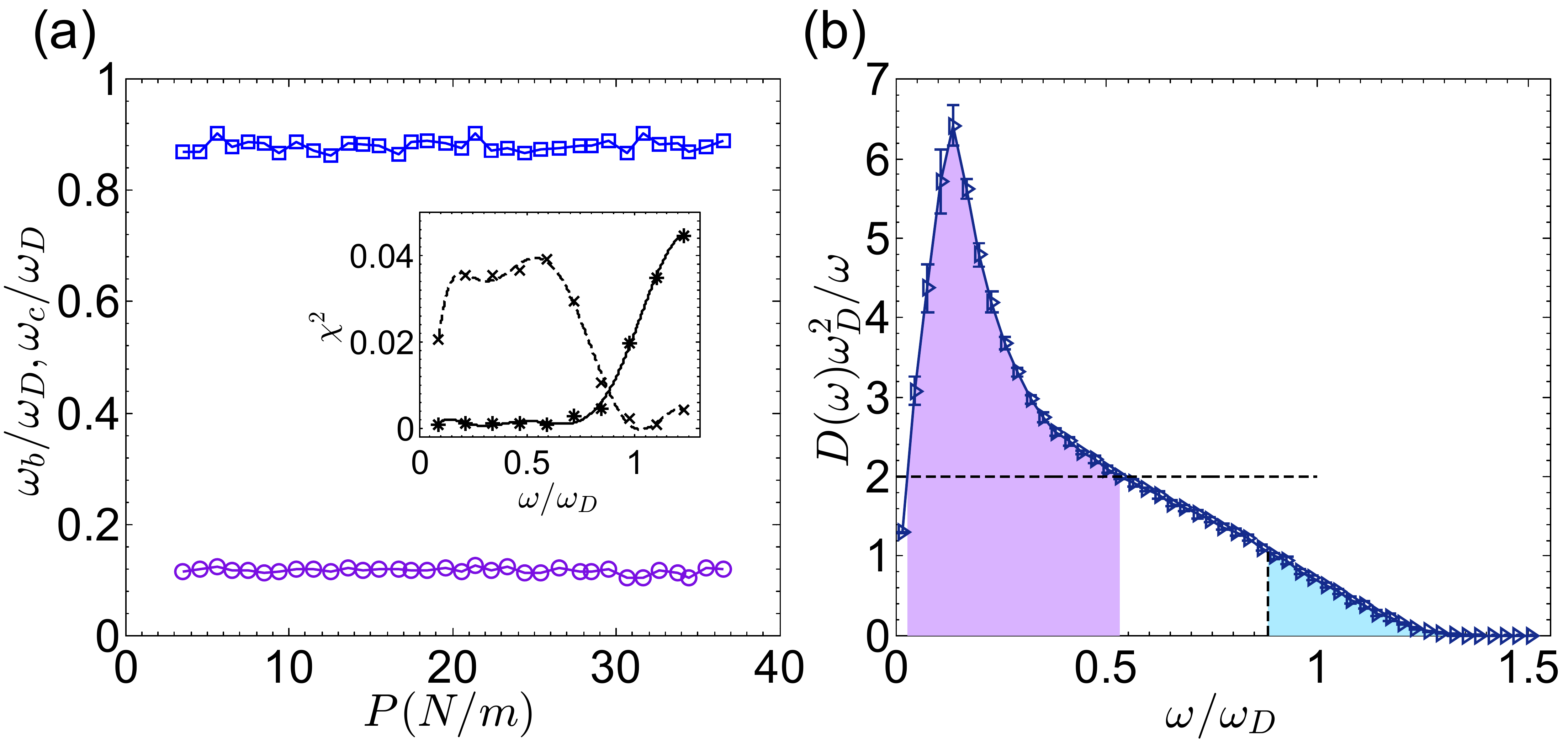}}
	\caption{\label{fig:figure2} 
	(a) Main panel: The boson-peak position $\omega_b/\omega_D$ (purple circles) and the $\omega_c/\omega_D$ (blue squares) versus pressure $P$, in which the results have been ensemble averaged over 10 realizations, and the error bars are within symbol sizes. 
	Inset: Chi-squared $\chi^2$ of the level-distance statistics fitted by two distributions of (1) GOE (asterisks) and (2) Poisson (crosses) versus the $\omega/\omega_D$. The two lines are smoothed curves to locate the critical point $\omega_c/\omega_D$ as their crossing point, which shows $\omega/\omega_D\approx0.88$ at pressure $P=26.5N/m$. 
\newline
	(b) Reduced density of states $g_\omega(\omega)/\omega$, re-scaled by the Debye frequency $\omega_D$ at a typical value of pressure $P=26.5N/m$. Here the lavender area denotes the boson-peak region, and the light blue denotes the Anderson-localization regime. Two dashed lines denote the Debye's model (a horizontal dashed line) and the $\omega_c/\omega_D$ (a vertical dashed line).}
\end{figure}

To identify the crossover frequency $\omega_c$, we quantitatively characterize deviations between data points and fitting curves using GOE and Poisson statistics by calculating the $\chi^2$ as a function of $\omega$, as shown in the inset of Fig.~\ref{fig:figure2}(a). Here, the $\chi^2=\sum (\tilde{y_i}-y_i)^2$, in which $\tilde{y_i}$ are the fitting values, and $y_i$ are the original data points. From the inset of Fig.~\ref{fig:figure2}(a), we see that when $\omega_0/\omega_D< \omega_c/\omega_D$, the distributions are better described by GOE statistics, while in the regime of $\omega_0/\omega_D \geq \omega_c/\omega_D$, distributions are better described by Poisson statistics. The crossover point occurs around $\omega_c/\omega_D =0.88\pm0.03$, which is nearly independent of pressure, as shown in the main panel (square symbols) of Fig.~\ref{fig:figure2}(a). The value of $\omega_c/\omega_D$ provides an estimate of the mobility edge. The determination of its precise value is   statistically more demanding \cite{2009-BJ.Huang-TM.Wu-PRE} and is beyond the present scope of this work. Also shown in the main panel of Fig.~\ref{fig:figure2}(a) are the positions of the boson peak, $\omega_b/\omega_D$, versus pressure $P$, which are independent of $P$, showing a constant value $\omega_b/\omega_D\approx 0.13$. Compared to the value of $\omega_b/\omega_D$, the $\omega_c/\omega_D$ is roughly 5.5 times larger, which indicates that the boson peak and Anderson localization are well separated. This separation is more obvious in Fig.~\ref{fig:figure2}(b), in which the two regimes, as depicted using two different colors -- lavender and light blue, are well separated, very simiuar to the three-dimensional case \cite{1998-W.Schirmacher-PRL}. 

\begin{figure}
\centerline{\includegraphics[trim=0cm 0cm 0cm 0cm, width=1\linewidth]{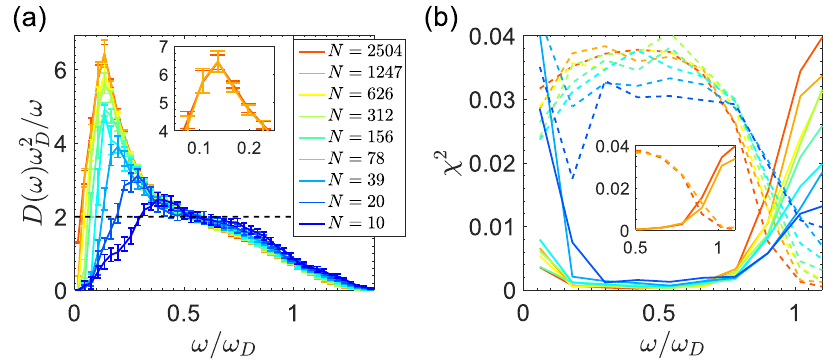}}
	\caption{\label{fig:figure3} (a) Reduced density of states $g_\omega(\omega)/\omega$ against $\omega/\omega_D$ of different system sizes at a typical value of pressure $P=26.5N/m$. Here, a horizontal dashed line represents the Debye's model. Inset: Amplified boson-peak regime of the two largest sizes. (b) Chi-squares $\chi^2$ vs the $\omega/\omega_D$ for different system sizes at a typical value of pressure $P=26.5N/m$. Here, solid lines represent the $\chi^2$ of GOE, while dashed lines are the $\chi^2$ of poisson distributions. Different colors indicate different sizes as defined in (a). Inset: Amplified $\chi^2$ curves of the two largest sizes. Note that results of a given size $N$ are ensemble averaged over $\sim \frac{2500}{N}\times9$ realizations.}
\end{figure}

To rule out any potential finite-size effect, we have done a systematic analysis of the finite-size effect by changing the system size $N$ from $N=10$ up to $N=2504$. The boson-peak position $\omega_b/\omega_D$ decreases when $N$ increases from $N=10$ up to $N=78$ and it quickly stabilizes to $\omega_b/\omega_D\approx0.13$ when $N\geq156$, as shown in Fig.~\ref{fig:figure3}(a). In addition, the upper limit of the boson-peak regime (i.e., the maximum frequency in the lavender color area of the boson peak regime) is a constant, nearly independent of $N$, as shown in Fig.~\ref{fig:figure3}(a). Moreover, shapes of the boson peak cease changing when $N\geq1247$ so that the lower limit of the boson-peak regime and the boson-peak height stabilize, as shown in Fig.~\ref{fig:figure3}(a). The stabilization of the boson-peak shape profile can be better viewed in the inset of Fig.~\ref{fig:figure3}(a), where the profiles of $N=1247$ and $N=2504$ collapse on top of each other. When the boson peak evolves with $N$, the Anderson-localization regime in the density of states also evolves, but its profile shape shows very little change as $N$ increases, as shown in Fig.~\ref{fig:figure3}(a). Meanwhile, $\omega_c/\omega_D$ changes more rapidly starting from $N=10$ and approaches a constant value of $\omega_c/\omega_D=0.88\pm0.03$ when $N\geq1247$, as shown in the main panel and the inset of Fig.~\ref{fig:figure3}(b). The above results suggest that the non-overlap between the boson peak and Anderson localization shown in Fig.~\ref{fig:figure2} can not be a finite-size effect.

\begin{figure}
\centerline{\includegraphics[trim=0cm 0cm 0cm 0cm, width=1\linewidth]{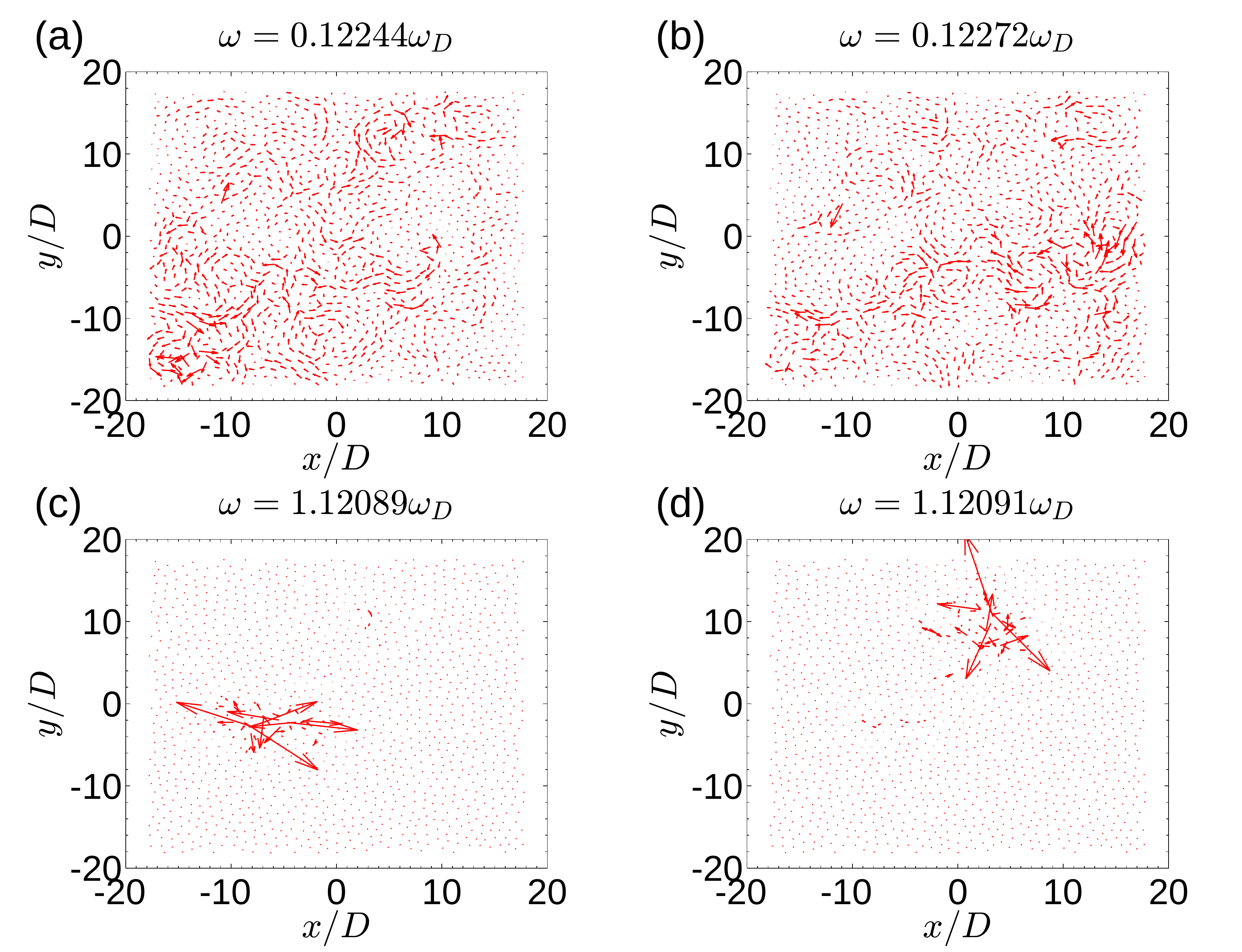}}
\caption{\label{fig:figure4} Comparison between spatial distributions of the modes within the Boson peak (a-b) and Anderson-localization regime (c-d). Here results are shown for a typical value of pressure $P=26.5N/m$, and results of other pressure are similar.}
\end{figure}

One of the nice things with the present experiment is that the spatial distributions of polarization vectors of modes can be directly visualized, as shown in Fig.~\ref{fig:figure4}.
In the Boson-peak regime, the spatial distributions of modes are extended and show strong heterogeneity, as shown in Fig.~\ref{fig:figure4}(a-b). In addition, the spatial distributions of the polarization vectors of two adjacent modes (i.e., $\lambda_i$ and $\lambda_{i+1}$) are not completely separated in space, but show different patterns, as shown in Fig.~\ref{fig:figure4}(a-b). In contrast, the spatial distributions of the modes in the high-frequency regime are completely localized, showing large displacements in a very limited region and essentially zero-amplitude displacements elsewhere in space, as shown in Fig~\ref{fig:figure4}(c-d). Further, the spatial distributions of two adjacent modes (i.e., $\lambda_i$ and $\lambda_{i+1}$) are completely separated in space, occupying markedly different spatial regions as shown in Fig~\ref{fig:figure4}(c-d). Moreover, the frequencies of two adjacent modes, and hence the values of $\lambda_i$ and $\lambda_{{i+1}}$ are much closer compared to those of the Boson-peak regime, as shown in Fig.~\ref{fig:figure4}. This indicates the absence of level repulsion, a further characteristic of localized states.

The finite-size analysis and the existence of a nonzero mobility edge $\omega_c$ seem to show clear contradiction to the theoretical prediction of complete Anderson localization in two dimension \cite{1985-Lee-RMP,1985-Kirkpatrick-PRB}. However, there is one possible caveat regarding the above finite-size analysis: according to the scaling theory \cite{1985-Lee-RMP}, or equally, the renormalization-group theory\cite{2018-Schirmacher-PRL,1983-John-PRB}, the localization length $\xi(\omega)$ grows extremely rapidly with decreasing frequency, which might go beyond any experimentally, or even numerically, accessible range of finite-size analysis. Recall that the localization length is given by \cite{1985-Lee-RMP}
\begin{equation}\label{eq:1}
	\xi(\omega)=D e^{\frac{\pi}{2}k\ell},
\end{equation}	
in which $D$ is the average particle diameter, $\ell$ is the mean-free path of wave, and $k=\omega/V_T$ with $V_T$ being the transverse wave velocity.	
This product $k\ell$ in the exponent can be expressed using the wave attenuation $\Gamma(\omega)$ as
\begin{equation}\label{eq:2}
k\ell=\omega/\Gamma(\omega) \sim \omega^{-2},
\end{equation}
with $\Gamma(\omega)\propto \omega^{d+1}$ (Rayleigh scattering) \cite{2010-Ganter-PRB,Wang2018-disentangling,2013-W.Schirmacher-ScientificRep,2009-G.Monaco-PNAS}, where $d$ is the dimensionality.
Using Eqs. (\ref{eq:1}) and (\ref{eq:2}), we can make a rough estimation of the value of $\xi(\omega)$ at $\omega_c$. 
From Figs. 4c and 4d we extract the localization length $\xi$ at
$\omega_1=1.1209\omega_D$ to be about $5D$. At such a high frequency the mean-free
path should be of the order of the disk diameter $D$. Taking Eqs. (\ref{eq:1}) and
(\ref{eq:2}) together we may effectively write
\begin{equation}\label{eq:3}
	\xi(\omega)=D\left(\frac{\scriptstyle \omega_1}{\scriptstyle \omega}\right)^3e^{A(\omega_D/\omega)^{-2}}
\end{equation}
Together with the above specified condition $\xi(\omega_1)=5D$ 
we obtain $A=2$. From Eq. (3) it follows 
$\xi(\omega_c)=26 D$, which is one-half of the system size.
Decreasing the frequency further leads to an immense increase of
the value of $\xi$, which may explain, why we find no dependence of
$\omega_c$ on the system size, as reported above.

Generally speaking, the relations (1), (2) and (3)
cause an essential singularity in the localization length,
which overrides the scaling towards localization. Therefore, in any finite two-dimensional system, be it very large, there will be a delocalization transition at a finite nonzero frequency. Even in the thermodynamic limit, the state at $\omega=0$ is delocalized\footnote{It is interesting to note that in electronic systems
with off-diagonal disorder the states at the band center (E=0) are
delocalized \cite{1998-Eilmes-EPJ,1998-Eilmes-PSS,2000-Biswas-PSS}}.

In summary, we have performed the first experimental study of level-distance statistics of normal-mode vibrations in densely packed two-dimensional packings of isotropically jammed photoelastic disks. We observe a clear wave delocalization for frequencies smaller than $\omega_c=0.88\pm0.03\omega_D$, which is in contradiction to theoretical predictions\cite{1985-Lee-RMP,1985-Kirkpatrick-PRB}. 
We believe that this observation is due to the fact that the localization length increases rapidly when frequency deceases causing a delocalization of waves at a nonzero frequency for any two-dimensional system of finite size.

We observe different level statistics and spatial distributions of polarization vectors in two separate regimes, namely at rather low frequency near the boson peak, where the modes are delocalized and at high frequeny, where the modes are Anderson-localized . The level-distance statistics in the Boson-peak regime is reasonably well described by the Wigner-Dyson GOE distribution, whereas the level-distance statistics of the Anderson-localization regime obeys the Poisson law. The spatial distributions of the polarization vectors of the modes in the Boson-peak regime are delocalized and exhibit strong heterogeneity, whereas those of the modes in the Anderson-localization regime are confined to small areas of a few particle diameters.

Our results are surprisingly similar to the situation in three dimensions
\cite{1998-W.Schirmacher-PRL}: delocalized modes near and above the boson peak and a mobility edge just below the Debye frequency leading to Anderson localization only at the upper end of the band of vibrational modes.

\begin{acknowledgments}
	This work is supported by the National Natural Science Foundation of China (NSFC) under (No.11774221 and No. 11974238). LZ also thanks the support by NSFC under No.11904410. 
\end{acknowledgments}
$^\dagger$equal contributions. $^{\ast}$(jiezhang2012@sjtu.edu.cn)


\end{document}